\documentclass[12pt,a4paper,english]{article}
\usepackage[latin9]{inputenc}
\usepackage{amsmath}
\usepackage{amssymb}
\usepackage{esint}

\makeatletter

\newcommand{\lyxaddress}[1]{
\par {\raggedright #1
\vspace{1.4em}
\noindent\par}
}

\usepackage{latexsym}

\usepackage{babel}

\makeatother

\usepackage{babel}
\begin{document}

\title{\begin{flushright}  {\small UWThPh-2017-18 \,}  \bigskip \bigskip \bigskip \end{flushright}Field
Space Entanglement Entropy,\\
Zero Modes and Lifshitz Models}

\author{Helmuth Huffel$^{\,\,1}$ and Gerald Kelnhofer$^{\,\,1}$ }

\maketitle

\lyxaddress{\begin{center}
$^{1}$$\,$ {\small Faculty of Physics, University of Vienna, Boltzmanngasse
5, A-1090 Vienna, Austria }\smallskip{}
 \\

\par\end{center}}
\begin{abstract}
\noindent The field space entanglement entropy of a quantum field
theory is obtained by integrating out a subset of its fields. We study
an interacting quantum field theory consisting of massless scalar
fields on a closed compact manifold $M$. To this model we associate
its Lifshitz dual model. The ground states of both models are invariant
under constant shifts. We interpret this invariance as gauge symmetry
and subject the models to proper gauge fixing. By applying the heat
kernel regularization one can show that the field space entanglement
entropies of the massless scalar field model and of its Lifshitz dual
are agreeing. 
\end{abstract}

\section{Introduction}

In quantum physics the entanglement entropy is a powerful and intriguing
observable (for reviews see \cite{Horodecki,Nielsen}) and has become
the subject of intensive investigation during the last decade. Entanglement
entropy provides valuable insight in condensed matter physics (for
reviews see \cite{Amico,Calabrese,Eisert,Laflorencie}), quantum field
theory (for reviews see \cite{Calabrese-1,Calabrese-2}) and black
hole theory (for reviews on holographic entanglement see \cite{Nishioka,Raamsdonk,Rangamani}).
\\
\\
Geometric entanglement entropy measures the entropy of a spatial subsystem
after tracing out the environment. Only rather few examples of field
theories are known, however, for which the entanglement entropy can
be computed exactly. In 1+1 dimensional conformal field theories the
replica technique pioneered by Cardy and Calabrese \cite{Calabrese 2004,Calabrese-2}
provides an indispensable computational tool. These methods have also
been extended to free field theories in higher dimensions \cite{Casini}
and to interacting theories using the AdS/CFT correspondence \cite{Nishioka,Rangamani}.
For a variational approximation to the entanglement entropy see \cite{Cotler}.\\
\\
The 2+1 dimensional quantum Lifshitz model is a continuum limit of
the dimer model \cite{Rokhsar,Moessner1,Moessner2} at the RK point
(for a review see \cite{Moessner3}) via a height function \cite{Henley,Kenyon}.
One of the remarkable properties of the model is that its ground state
wave functional is exactly known and that it has two-dimensional conformal
invariance. Therefore techniques of conformal field theory are applicable
and analytic calculations become possible; \cite{Fradkin,Hsu,Stephan,Fradkin 2009,Oshikawa,Zaletel,Zhou}
demonstrated that the entanglement entropy for various geometries
could be expressed and calculated in terms of the free energy of the
associated conformal field theory. \\
\\
Correlation functions of a $n$ dimensional interacting quantum field
theory are coinciding with the large equal-time correlation functions
of an associated ``dual'' Lifshitz model in $(n+1)$ dimensions.
This specific relationship has been one of the central issues in stochastic
quantization (\cite{Parisi}, for reviews see \cite{Damgaard,namiki})
and can be summarized as follows: One starts from a Euclidean field
theory model in $n$ dimensions and introduces an ergodic stochastic
process possessing the $n$ dimensional Euclidean path integral measure
as its unique equilibrium measure. The stochastic process evolves
in an additional time and is defined through a Fokker-Planck equation.
Its drift force is taken to be the negative variation of the $n$
dimensional Euclidean action with respect to the fields. Employing
the Feynman-Kac formula for the positive semidefinite Fokker-Planck
Hamiltonian one is lead to a functional integral of a field theory
model - called dual Lifshitz model - in $(n+1)$ dimensions. The partition
function of the original $n$ dimensional model is expressed as the
norm square of the ground state of the new $(n+1)$ dimensional model.
\\
\\
The prototypical example of the duality of field theory models due
to the stochastic quantization scheme is the duality of a free scalar
field and the Lifshitz model. In recent years novel interpretations,
applications and generalizations of this duality were given (for reviews
see \cite{Ardonne,Dijkgraaf}), Lifshitz-type models of gauge and
gravity theories \cite{Horava1,Horava2} attracted great interest.
\\
\\
In this work we are interested in the field space entanglement entropy
of quantum field theories, which is obtained by integrating out a
subset of its fields. Various measures for entanglement as well as
the formulation of holographic entanglement entropy have been investigated
in this context \cite{Mollabashi1,Mollabashi2,Taylor}. We address
the question to which extent the field space entanglement entropies
of quantum field theories and their Lifshitz duals are agreeing.\\
\\
Specifically we compare the field space entanglement entropy of the
dual Lifshitz model associated to a model of two interacting massless
scalar fields on a compact and closed manifold $M$ with the field
space entanglement entropy of the same model, yet defined on ${\bf {\bf \mathbb{R}}}\times M$.
The ground states of these two models are Gaussian wave functionals invariant
under constant shifts and thus are not normalizable. This is related
to the existence of zero modes of the Laplacian on compact and closed
manifolds.  In fact, zero modes represent gauge degrees of freedom,
see \cite{folacci} for a gauge theory interpretation of zero modes
in the discussion of the quantization of a massless scalar field on
${\bf \mathbb{S}}^{4}$. The - non gauge theory - role of zero modes
for the geometrical entanglement entropy is discussed in \cite{Mallayya,Zhou,yazdi,klich}
and for the field space entanglement entropy in \cite{Mollabashi1,Mollabashi3}.
\\
\\
It is our focus to carefully handle zero modes in field space entanglement
entropy. Instead of regulating  ground states by ad hoc procedures
for the zero modes, we interpret these invariances as gauge symmetries
and subject the models to proper gauge fixing \cite{annals2,Kelnhofer}.\\
\\
The presence of gauge symmetries adds a subtle component to the construction
of the dual Lifshitz model and  its ground state, which we are going
to review in the following lines. Remind the origin of Lifshitz dual
models, which is lying in the Fokker-Planck formulation of stochastic
quantization. Parisi and Wu \cite{Parisi} proposed stochastic quantization
without gauge fixing terms. Since the pure action of the gauge model
remains invariant under gauge transformations, the associated drift
force acts orthogonally to the gauge orbits. As a consequence unbounded
diffusion along the orbits takes place and a Fokker Planck formulation
is not possible. In the approach of Zwanziger \cite{zwanziger} the
drift force of the stochastic process is modified by the addition
of a gauge fixing force tangent to the gauge orbits. This provides
damping for the gauge modes' diffusion along the orbits leaving unchanged
all gauge invariant expectation values. Zwanziger's gauge fixing 
force, however, intrinsically is non-conservative and generally cannot
be accomodated in an action formalism required for the Lifshitz model
construction. Eventually, the stochastic quantization approach could
be generalized \cite{annals1,annals2} by specifying a drift force
which not only has tangential components along the gauge orbits but
where, subsequently, also the Wiener process itself is modified. Expectation
values of gauge invariant observables again remain untouched. With
\cite{annals1,annals2} a well defined Fokker-Planck formulation with
normalizable ground state is obtained and  the corresponding Lifshitz
model can be derived in a consistent way. As a consequence the issue
of comparing field space entanglement entropies of quantum field theories
and of their Lifshitz duals can be addressed.\\
\\
For massless scalar fields the naive set up of a dual Lifshitz model
fails in the first step due to the gauge invariance of the model.
A consistent formulation is presented in section 2 after the proper
discussion of gauge symmetries and gauge fixing.\\
\\
In section 3 we construct the regularized ground state associated
to the massless scalar field model in the Hilbert space of the entire
system, whose inner product is given by integration over all scalar
fields. In order to obtain a normalizable ground state, we modify
the Lagrangian of the massless scalar field model by adding the same
gauge fixing term as in section 2.\\
\\
By applying the heat kernel regularization we can show in section
4 that the field space entanglement entropies of the massless scalar
field model and of its Lifshitz dual are agreeing.

\section{Regularized ground state of the dual Lifshitz model}

Let $M$ be a compact $n$ dimensional closed manifold and consider
the action functional of coupled massless real-valued scalar fields
$\phi_{1}(x),\phi_{2}(x)$ on $M$ 
\begin{equation}
S^{M}(\phi_{1},\phi_{2})=\frac{1}{2}\int_{M}dvol(x)\,[\phi_{1}\triangle^{M}\phi_{1}+\phi_{2}\triangle^{M}\phi_{2}+\lambda\phi_{1}\triangle^{M}\phi_{2}]\label{eq:gauge}
\end{equation}
where $\lambda$ is a coupling parameter. Let\textsc{ ${\cal F}$}
denote the configuration space of the fields. To this model in a first
step we associate a dual Lifshitz model \cite{Parisi,Damgaard,namiki,Ardonne,Dijkgraaf},
such that its ground state, called Lifshitz ground state, is given
by
\begin{equation}
\tilde{\Psi}_{0}^{Lif}(\phi_{1},\phi_{2})=N\, e^{-\frac{1}{2}S^{M}(\phi_{1},\phi_{2})},\label{eq:lifshitz ground}
\end{equation}
where $N$ should be a normalization constant. It is well known, however,
that $e^{-\frac{1}{2}S^{M}(\phi_{1},\phi_{2})}$ is not normalizable
due to the invariance of $S^{M}(\phi_{1},\phi_{2})$ under constant
shifts $(\phi_{1},\phi_{2})\longmapsto(\phi_{1}+c_{1},\phi_{2}+c_{2})$.
We interpret this invariance of the Lifshitz ground state as being
associated to the gauge transformation
\begin{equation}
{\cal F}\times{\bf {\bf \mathbb{R}}^{2}}\longmapsto{\cal F}\qquad(\phi_{1},\phi_{2},c_{1},c_{2})\longmapsto(\phi_{1}+c_{1},\phi_{2}+c_{2}).\label{eq:gauge transformation FxR2}
\end{equation}
Hence the quotient ${\cal F}\diagup{\bf \mathbb{R}}^{2}$ is the true
configuration space of the coupled system giving rise to the trivializable
principal ${\bf {\bf \mathbb{R}}^{2}}$-bundle ${\cal F}\longmapsto{\cal F}\diagup{\bf {\bf \mathbb{R}}^{2}}$,
with trivialization
\begin{equation}
\omega(\phi_{1},\phi_{2})=\frac{1}{V_{M}}(\int_{M}dvol(x)\,\phi_{1},\int_{M}dvol(x)\,\phi_{2})=(\Pi(\phi_{1}),\Pi(\phi_{2})),
\end{equation}
where $V_{M}=\int_{M}dvol(x)$ is the volume of $M$, and $\Pi$ is
the projector onto the constant parts of the scalar fields. In fact,
the space ${\cal F}\diagup{\bf \mathbb{R}}^{2}$ of the physical degrees
of freedom can be identified with the space of pairs of non-constant
scalar fields. \\
\\
Evidently $\omega$ fulfills $ $$\omega(\phi_{1}+c_{1},\phi_{2}+c_{2})=\omega(\phi_{1},\phi_{2})+(c_{1},c_{2})$
(see \cite{annals2,Kelnhofer} for further details, and \cite{folacci}
for a BRST approach). \\
\\
As outlined in the previous section we interpret the Lifshitz model
as being associated to the Focker-Planck Hamiltonian arising in the
stochastic quantization of the gauge model of massless scalar fields
(\ref{eq:gauge}). Following the general procedure in \cite{annals1,annals2,Kelnhofer}
the underlying stochastic process has to be adapted judiciously, according
to the choice of a gauge fixing function $(w^{*}S_{gf})(\phi_{1},\phi_{2})$.
The gauge invariant action $S^{M}(\phi_{1},\phi_{2})$ then is replaced
by the gauge fixed total action $S_{tot}^{M}(\phi_{1},\phi_{2})$,
which is to be used for the proper definition of the Lifshitz model.
\\
\\
For the gauge fixing function we choose
\begin{equation}
S_{gf}(c_{1},c_{2})=\frac{\mu}{2}(c_{1}^{2}+c_{2}^{2}+\lambda c_{1}c_{2}),
\end{equation}
which for $\mu>0$ and $-2<\lambda<2$ is normalizable. Thus we arrive
at a gauge fixed total action
\begin{equation}
\begin{array}{ccc}
S_{tot}^{M}(\phi_{1},\phi_{2}) & = & S^{M}(\phi_{1},\phi_{2})+(w^{*}S_{gf})(\phi_{1},\phi_{2})\\
 & = & \frac{1}{2}\int_{M}dvol(x)\,[\phi_{1}D^{M}\phi_{1}+\phi_{2}D^{M}\phi_{2}+\lambda\phi_{1}D^{M}\phi_{2}],
\end{array}\label{total lifshitz action}
\end{equation}
where we introduced $D^{M}=\triangle^{M}+\frac{\mu}{V_{M}}\Pi$. The
Laplacian $\triangle^{M}$ has discrete spectrum 
\begin{equation}
0=\nu_{0}(\triangle^{M})<\nu_{1}(\triangle^{M})\leq\nu_{2}(\triangle^{M})\leq...\;\longrightarrow\infty,
\end{equation}
where each eigenvalue appears the same number of times as its multiplicity.
Let $\chi_{\alpha}^{M}$ denote the orthonormal basis of eigenfunctions
of $\triangle^{M}$ such that
\begin{equation}
\triangle^{M}\chi_{\alpha}^{M}=\nu_{\alpha}(\triangle^{M})\chi_{\alpha}^{M}.
\end{equation}
Then $\chi_{\alpha}^{M}$ are also eigenfunctions of $D^{M}$ with
the eigenvalues
\begin{equation}
D^{M}\chi_{\alpha}^{M}=\nu_{\alpha}(\triangle^{M})\chi_{\alpha}^{M}+\frac{\mu}{V_{M}}\delta_{\alpha0}\chi_{\alpha}^{M}.
\end{equation}
The strictly positive spectrum of $D^{M}$ reads
\begin{equation}
Spec(D^{M})=\left\{ \nu_{\alpha}(\triangle^{M})_{\vert\alpha\neq0},\frac{\mu}{V_{M}}\right\} .
\end{equation}
With respect to the orthonormal basis $\left\{ \chi_{\alpha}^{M}\right\} $
we rewrite
\begin{equation}
\phi_{1}=\sum_{\alpha}\phi_{1,\alpha}\chi_{\alpha}^{M},\qquad\phi_{2}=\sum_{\alpha}\phi_{2,\alpha}\chi_{\alpha}^{M}.
\end{equation}
Thus instead of (\ref{eq:lifshitz ground}) we define the regularized
Lifshitz ground state with respect to the gauge fixed total action
(\ref{total lifshitz action}) by
\begin{equation}
\Psi_{0}^{Lif}(\phi_{1},\phi_{2})=\prod_{\alpha\neq0}\Psi_{0}^{Lif}(\phi_{1,\alpha},\phi_{2,\alpha})\Psi_{0}^{Lif}(\phi_{1,0},\phi_{2,0}),
\end{equation}
where
\begin{equation}
\Psi_{0}^{Lif}(\phi_{1,\alpha},\phi_{2,\alpha})=Ne^{-\frac{1}{4}\nu_{\alpha}(\triangle^{M})\left(\phi_{1,\alpha}^{2}+\phi_{2,\alpha}^{2}+\lambda\phi_{1,\alpha}\phi_{2,\alpha}\right)},
\end{equation}
as well as
\begin{equation}
\Psi_{0}^{Lif}(\phi_{1,0},\phi_{2,0})=Ne^{-\frac{1}{4}\frac{\mu}{V_{M}}\left(\phi_{1,0}^{2}+\phi_{2,0}^{2}+\lambda\phi_{1,0}\phi_{2,0}\right)}.
\end{equation}
 
\[
\]

\section{Regularized ground state of the massless scalar field model}

As a next step we construct the ground state associated to the massless
scalar field model $S^{M}(\phi_{1},\phi_{2})$. It will be convenient
to define this model on ${\bf {\bf \mathbb{R}}}\times M$ by the action
functional
\begin{equation}
S^{\mathbb{R}\times M}=\int_{R}dt\, L^{M}(\phi_{1},\phi_{2},\dot{\phi}_{1},\dot{\phi}_{2})
\end{equation}
 with
\begin{equation}
L^{M}(\phi_{1},\phi_{2},\dot{\phi}_{1},\dot{\phi}_{2})=\frac{1}{2}\int_{M}dvol(x)\,[\dot{\phi}_{1}{}^{2}+\dot{\phi}_{2}{}^{2}+\lambda\dot{\phi}_{1}\dot{\phi}_{2}]+S^{M}(\phi_{1},\phi_{2}).\label{lagrangian}
\end{equation}
 This system, like the dual Lifshitz model before, admits the gauge
symmetry (\ref{eq:gauge transformation FxR2}) under constant shifts.
A straightforward Hamiltonian analysis following the Dirac constraint
quantization approach \cite{Dirac1,Dirac2} shows that the corresponding
physical states are wave functions $\tilde{\Psi}(\phi_{1},\phi_{2})$
which are constant under the field transformations (\ref{eq:gauge transformation FxR2}).
The ground state is
\begin{equation}
\tilde{\Psi}_{0}(\phi_{1},\phi_{2})=N\, e^{-\tilde{S}(\phi_{1},\phi_{2})}.\label{ground state}
\end{equation}
where 
\begin{equation}
\tilde{S}(\phi_{1},\phi_{2})=\frac{1}{2}\int_{M}dvol(x)\,[\phi_{1}\left(\triangle^{M}\right)^{\frac{1}{2}}\phi_{1}+\phi_{2}\left(\triangle^{M}\right)^{\frac{1}{2}}\phi_{2}+\lambda\phi_{1}\left(\triangle^{M}\right)^{\frac{1}{2}}\phi_{2}].\label{eq:groundstate2}
\end{equation}
Since the ground state is constant along the gauge orbits, it is not
normalizable with respect to the Hilbert space of the entire system
whose inner product is given by integration over all scalar fields.
In order to obtain a normalizable ground state, we modify the Lagrangian
(\ref{lagrangian}) by adding the same gauge fixing term $(w^{*}S_{gf})(\phi_{1},\phi_{2})$
as before. Hence the system remains unchanged along the physical degrees
of freedom and we get
\begin{equation}
L_{tot}^{M}(\phi_{1},\phi_{2},\dot{\phi}_{1},\dot{\phi}_{2})=\frac{1}{2}\int_{M}dvol(x)\,[\dot{\phi}_{1}{}^{2}+\dot{\phi}_{1}{}^{2}+\lambda\dot{\phi}_{1}\dot{\phi}_{2}]+S_{tot}^{M}(\phi_{1},\phi_{2}).
\end{equation}
The mode expansion on $M$ leads to the regularized ground state
\begin{equation}
\Psi_{0}(\phi_{1},\phi_{2})=\prod_{\alpha\neq0}\Psi_{0}(\phi_{1,\alpha},\phi_{2,\alpha})\Psi_{0}(\phi_{1,0},\phi_{2,0}),
\end{equation}
where
\begin{equation}
\Psi_{0}(\phi_{1,\alpha},\phi_{2,\alpha})=Ne^{-\frac{1}{2}\left(\nu_{\alpha}(\triangle^{M})\right)^{\frac{1}{2}}\left(\phi_{1,\alpha}^{2}+\phi_{2,\alpha}^{2}+\lambda\phi_{1,\alpha}\phi_{2,\alpha}\right)},
\end{equation}
as well as
\begin{equation}
\Psi_{0}(\phi_{1,0},\phi_{2,0})=Ne^{-\frac{1}{2}\left(\frac{\mu}{V_{M}}\right)^{\frac{1}{2}}\left(\phi_{1,0}^{2}+\phi_{2,0}^{2}+\lambda\phi_{1,0}\phi_{2,0}\right)}.
\end{equation}
The modified ground state agrees, when restricted to the physical
degrees of freedom, with the ground state (\ref{ground state},\ref{eq:groundstate2})
of the original system. Its essential effect is to provide an integrable
contribution along the gauge orbits and thus, like in the dual Lifshitz
model, gives a well defined state in the full Hilbert space.

\section{Entanglement entropy}

So far we provided the mode decompositions of the regularized ground
states for the two considered scalar field models. Each mode contributed
in a similar way
\begin{equation}
\Psi_{\alpha}(\phi_{1,\alpha},\phi_{2,\alpha})=Ne^{-c_{\alpha}\left(\phi_{1,\alpha}^{2}+\phi_{2,\alpha}^{2}+\lambda\phi_{1,\alpha}\phi_{2,\alpha}\right)},
\end{equation}
where the positive constants $c_{\alpha}$ can be found in the equations
of the previous sections (their values are irrelevant for the calculation
of the entanglement entropy, however). The ultimate calculation of
the field space entanglement entropy is quickly done: A simple analysis
along the lines of \cite{bombelli,Srednicki,callan,Mollabashi1} tells
us that each mode's contribution is governed by the eigenvalues $\left\{ \lambda_{1},\lambda_{2}\right\} $
of the quadratic form $c_{\alpha}\left(\phi_{1,\alpha}^{2}+\phi_{2,\alpha}^{2}+\lambda\phi_{1,\alpha}\phi_{2,\alpha}\right)$.
These eigenvalues are simply $\left\{ \lambda_{1,\alpha},\lambda_{2,\alpha}\right\} =\left\{ c_{\alpha}\frac{2-\lambda}{2},c_{\alpha}\frac{2+\lambda}{2}\right\} $
and one calculates the auxiliary quantity
\begin{equation}
\xi_{\alpha}=\left(\frac{\sqrt{\lambda_{1,\alpha}}-\sqrt{\lambda_{2,\alpha}}}{\sqrt{\lambda_{1,\alpha}}+\sqrt{\lambda_{2,\alpha}}}\right)^{2}=-\frac{-8+\lambda^{2}+4\sqrt{4-\lambda^{2}}}{\lambda^{2}},
\end{equation}
which is independent of $\alpha$, independent of $c_{\alpha}$, independent
of the gauge fixing parameter $\mu$ and is equal for the scalar field
model and its Lifshitz dual. Finally one obtains for each mode - also
the zero mode - an equal contribution $s$ to the field space entanglement
entropy
\begin{equation}
s_{\alpha}=-log(1-\xi_{\alpha})-\frac{\xi_{\alpha}}{1-\xi_{\alpha}}log(\xi_{\alpha})=s.
\end{equation}
 We perform the heat kernel regularized sum over all the contributions
of the modes
\begin{equation}
S_{reg}(t)=\sum_{\alpha=0}^{\infty}s_{\alpha}e^{-t\,\nu_{\alpha}(\triangle^{M})}=s\sum_{\alpha=0}^{\infty}e^{-t\,\nu_{\alpha}(\triangle^{M})}.
\end{equation}
Using the asymptotic expansion for the heat kernel of the Laplacian,
we finally obtain for the regularized entanglement entropy 
\begin{equation}
S_{reg}(t)\simeq s\,\sum_{k=0}^{\infty}a_{k}(\triangle^{M})t^{\frac{k-n}{2}},\qquad t\searrow0
\end{equation}
where $a_{k}(\triangle^{M})$ are the Seeley coefficients \cite{Gilkey}
of the Laplacian on $M$. Since 
\begin{equation}
a_{0}(\triangle^{M})=\left(4\pi\right)^{-\frac{n}{2}}\, V_{M}
\end{equation}
one thus arrives at the field space entanglement being proportional
to the volume of $M$ \cite{Mollabashi1,Mollabashi2,Taylor}. \\
\\
Concluding, we have shown that the field space entanglement entropies
of the massless scalar field model and of its Lifshitz dual are agreeing.

\section{Outlook}

It seems interesting to adapt our present calculation for the case
of self interacting scalar fields along the lines of \cite{Cotler}.
\\
\\
In a forthcoming publication, we will present calculations of the
field space entanglement entropy in the case of a gauge model with
Gribov ambiguities \cite{paper}.

\section*{Acknowledgments }

We thank Harald Grosse, Daniel Grumiller and Beatrix Hiesmayr for
valuable discussions.

\end{document}